\documentclass{ws-procs9x6}

\usepackage{subfig}
\usepackage[english]{babel}

\begin{document}

\title{Signal Attenuation Curve for Different Surface Detector Arrays}

\author{J. VICHA$^*$, P. TRAVNICEK$^*$, D. NOSEK$^+$ and J. EBR$^*$}

\address{$^*$Institute of Physics of Academy of Sciences, Prague, Czech Republic}

\address{$^+$Charles University, Faculty of Mathematics and Physics, Prague, Czech Republic}
E-mail: vicha@fzu.cz\\

\begin{abstract}
Modern cosmic ray experiments consisting of large array of particle detectors measure the signals of electromagnetic or muon components or their combination. The correction for an amount of atmosphere passed is applied to the surface detector signal before its conversion to the shower energy. Either Monte Carlo based approach assuming certain composition of primaries or indirect estimation using real data and assuming isotropy of arrival directions can be used. Toy surface arrays of different sensitivities to electromagnetic and muon components are assumed in MC simulations to study effects imposed on attenuation curves for varying composition or possible high energy anisotropy. The possible sensitivity of the attenuation curve to the mass composition is also tested for different array types focusing on a future apparatus that can separate muon and electromagnetic component signals.
\end{abstract}


\bodymatter

\section{Introduction}
Two largest modern experiments use scintillator detectors (Telescope Array) or water Cherenkov detectors (Pierre Auger Observatory) at ground for studying cosmic ray showers of ultra-high energies (higher than $10^{18}$~eV). Both experiments are located at the approximately same altitude (around 1400~m~a.s.l. equivalent to 880~g/cm$^{2}$ of atmospheric depth). Thin scintillator detectors are dominantly sensitive only to the electromagnetic (EM) component, while in water Cherenkov detectors the signal is produced by EM particles and muons (EM+$\mu$) as well. In any case, the signal ($S$) of the surface detector array has to be corrected for an attenuation of shower size with respect to the amount of air penetrated before reaching the detector. In principle, there are two options how to correct for this effect: assume certain composition of primaries and use MC simulations or assume isotropy of arrival directions above certain energy and use measured data. 

Telescope Array uses the so called look-up table from Monte Carlo (MC) simulations providing the relation between the signal size, zenith angle ($\Theta$) and the shower energy~\cite{TA:enespec}. Only proton primaries are considered. At the Pierre Auger Observatory, the so called Constant Intensity Cut (CIC) method \cite{Hersil:cicmet} is applied to the measured data \cite{AUGER:enespec} providing a relationship between the size of the signal and the $cos^2(\Theta)$ at a given intensity (energy) cut. In the next step, for each shower the signal at the reference angle ($S_{Ref}$) is calculated using normalized CIC curve. The $S_{Ref}$ value is then related to the shower energy measured by the fluorescence detector. The CIC curve is studied as a function of energy (intensity cut). Since no substantial deviations in the CIC curve shape are found, just one normalized curve is finally used for all the showers.

For our studies we use Toy MC in combination with an output from simulations produced by CORSIKA ver.~7.37~\cite{CORSIKA}. Based on very rough assumptions of the detector response we first calculate examples of the signal attenuation curves for proton and iron induced showers from the CORSIKA simulations. We assumed responses for EM, EM+$\mu$ and $\mu$ types of observatories. These curves with an assumed size of signal fluctuations and energy independence then serve as an input for Toy MC to generate a large number of events that are reconstructed by both MC-like and CIC-like approaches.  

In our previous study \cite{ICRC} we showed what happens to reconstructed energies if the primary particles are of a mixed composition from protons and iron nuclei. Both possibilities - the application of the MC attenuation curve and the CIC approach were investigated separately for EM type as well as EM+$\mu$ sensitive observatory. Since the CIC method is based on the assumption of the uniform distribution of events in $cos^2(\Theta)$, we also tried to estimate the influence of presence of a source at the highest energies violating to some extend this uniform distribution. 

Briefly summarizing, conclusions are as follows: MC approach applied to mixed composition yields to zenith angle bias on reconstructed energies. CIC approach provides stable energy reconstruction for any composition mixture. CIC approach is valid even for presence of very strong sources at the highest energies. 

In this study we rather consider two types of surface detectors at a single observatory. A sensitivity to the mass composition of primaries utilizing different responses to iron nuclei and protons is shown using the same approach with CORSIKA shower characteristics as inputs into Toy MC. More details about our method can be found in our previous study.

 \begin{figure}[h!]
  \centering
\includegraphics[width=0.95\textwidth]{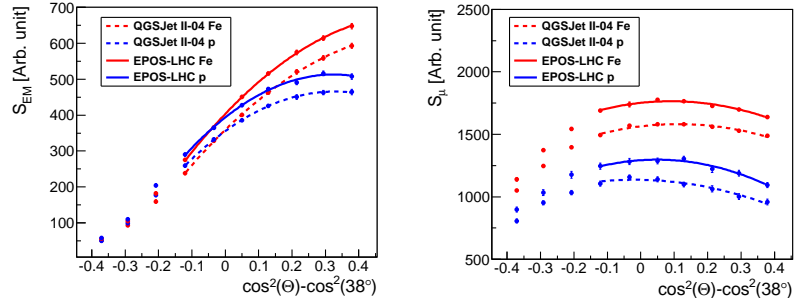}
  \caption{Averaged signals fitted with quadratic function of $cos^{2}(\Theta)$ for EM detector (left panel) and $\mu$ detector (right panel) in the range [$0^{\circ},45^{\circ}$].}
  \label{AbsoluteSignals}
 \end{figure}

\section{Two Surface Detectors at Single Observatory}

\begin{figure}[h!]
  \centering
  \subfloat{\includegraphics[width=0.47\textwidth]{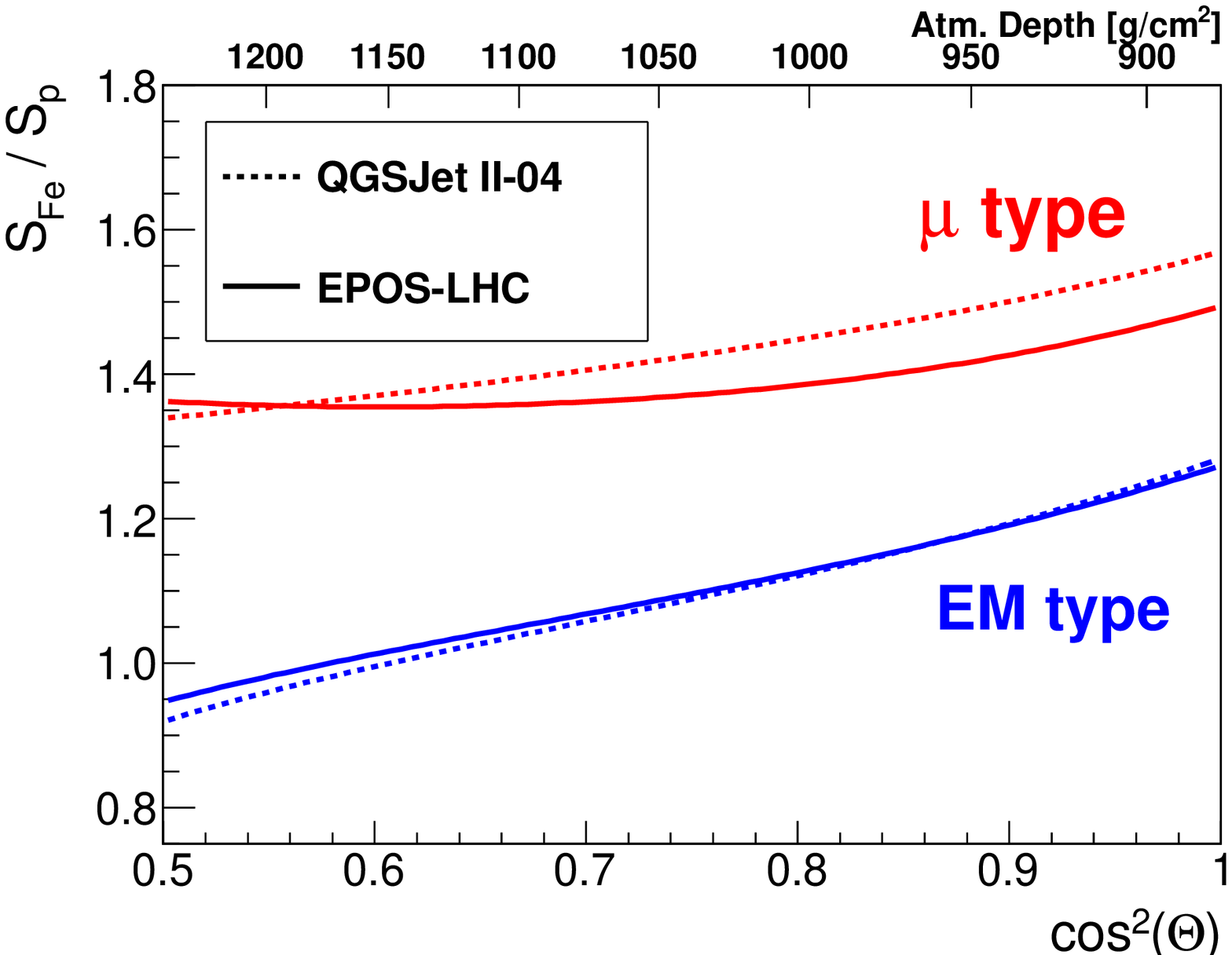}}
  \subfloat{\includegraphics[width=0.53\textwidth]{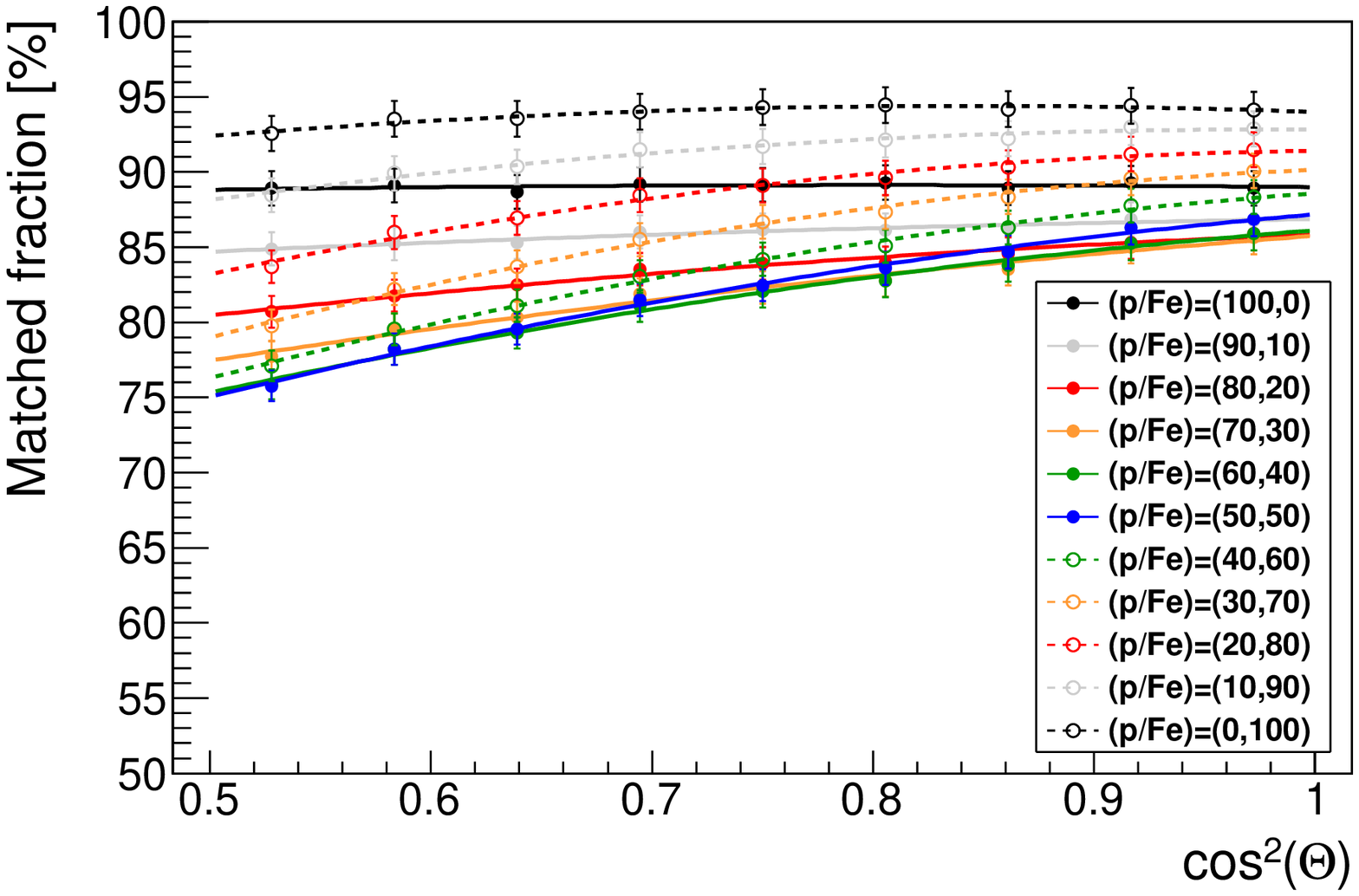}}
  \caption{Left panel: Ratios of detector responses of iron nuclei to protons for EM detector (blue) and $\mu$ detector (red). Right panel: Relative number of the same events in EM and $\mu$ detector with signal sizes above a signal size corresponding to $\sim$10~EeV.}
  \label{ResponsesFeP}
 \end{figure}

We assumed two different detector responses at the same observatory located 1400~m~a.s.l. Response of the first detectors is proportional to the density of EM particles at ground (threshold energy~1~MeV) and the response of the second type of detectors to the muon density (threshold energy~50~MeV) produced by CORSIKA. Lateral distribution function at 1000~m from the shower core was chosen as a reference signal ($S_{\mu}$,~$S_{EM}$). Zenith angle dependence of averaged signals at 10$^{19}$~eV of both detectors is shown in Fig.~\ref{AbsoluteSignals} for p, Fe primaries and two hadronic interaction models: QGSJet~II-04~\cite{QGSJETII04} and EPOS-LHC~\cite{epos}\cite{epos2}. As common, FLUKA model~\cite{FLUKA} was used for low energy interactions. For one zenith angle, one primary type and one hadronic interaction model about 60 CORSIKA showers were simulated. Corresponding ratio of detector responses of iron nuclei to protons ($S_{Fe}/S_{p}$) depending on $cos^{2}(\Theta)$ is plotted on the left panel of Fig.~\ref{ResponsesFeP}. Note that $S_{Fe}/S_{p}$ for $\mu$ detector is always above~1 and for EM detector $S_{Fe}/S_{p}$ decreases even below~1 at $cos^{2}(\Theta)=0.5$. 

The relative attenuations for protons and iron nuclei together with the ratio of responses for QGSJet~II-04 were included in the Toy MC generating shower energies in the range $[10^{18.5},10^{20}]$~eV with spectral index 2.7 and the GZK feature at the end of the spectrum. The zenith angle was distributed uniformly from 0 to 60 degrees. In this way, we obtained $10^{6}$ triplets of $S_{\mu}$, $S_{EM}$ and $\Theta$ for different mass compositions. Only events with zenith angle less than 45$^{\circ}$ were used in the further analysis to fulfil the full trigger efficiency of EM detectors in current arrays.

\section{Sensitivity to Mass Composition}

\begin{figure}[t]
  \centering
  \subfloat{\includegraphics[width=0.5\textwidth]{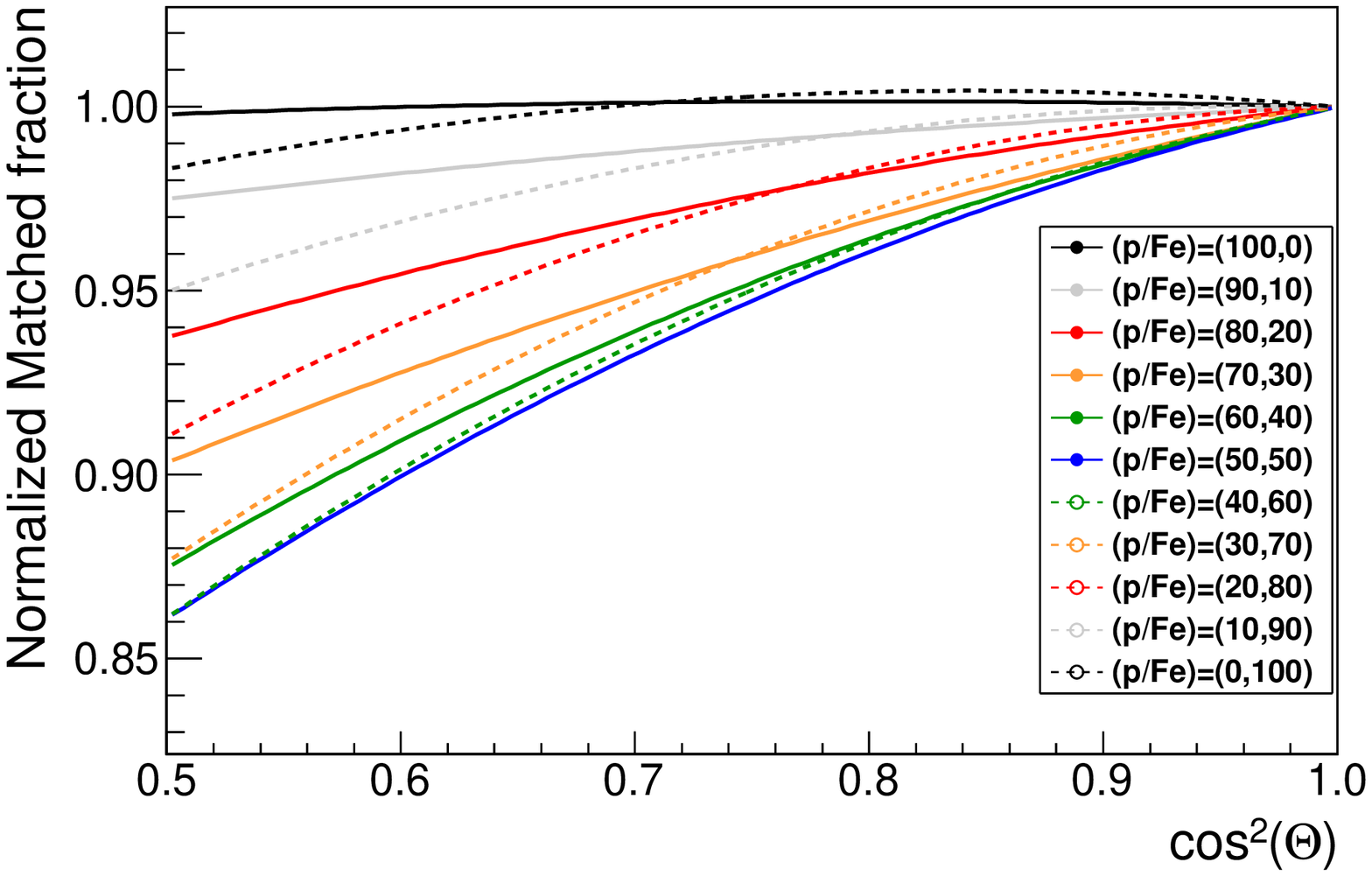}}
  \subfloat{\includegraphics[width=0.5\textwidth]{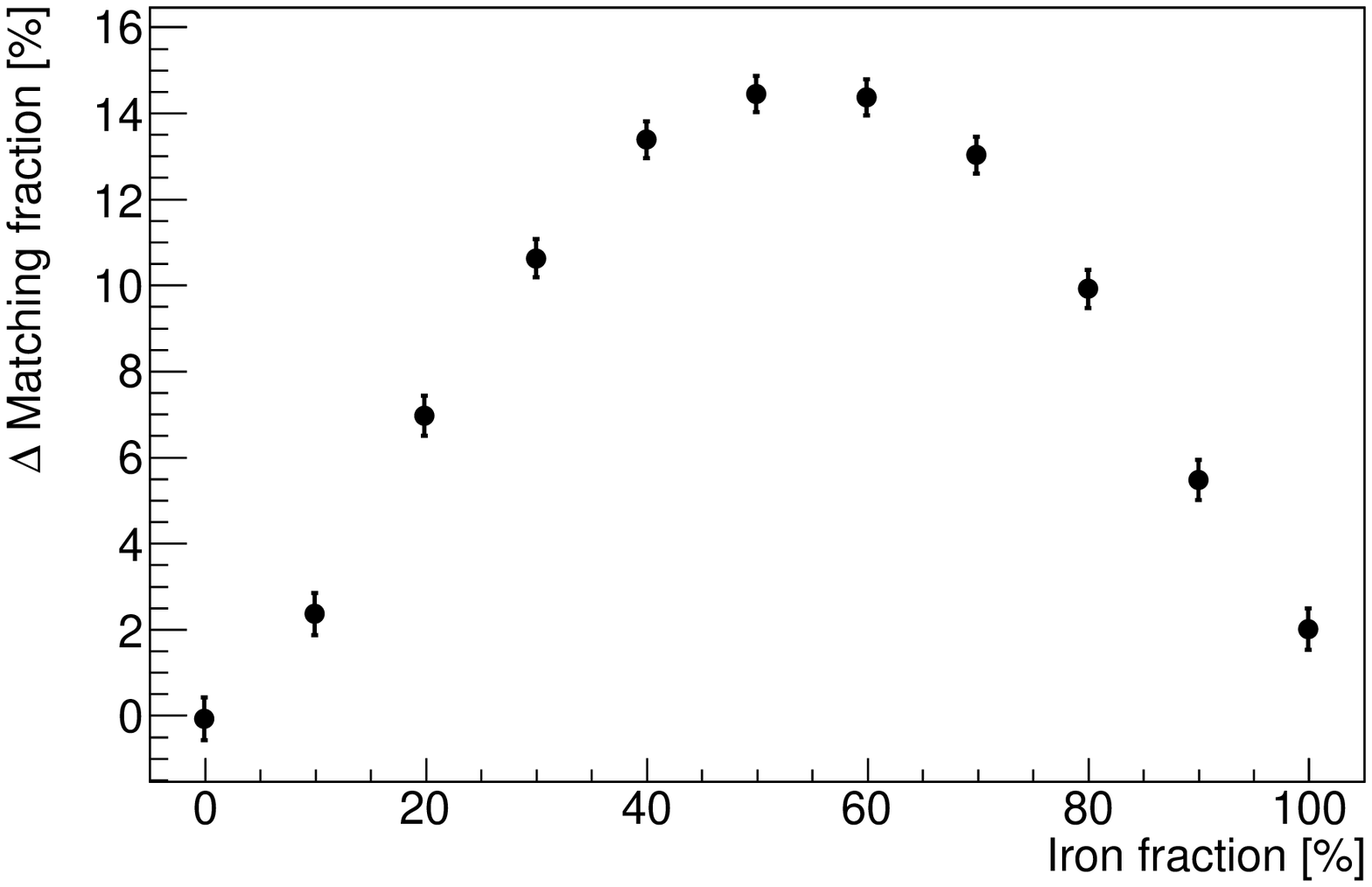}}
  \caption{Left~panel:~The fitted quadratic lines from the plot on the left panel of Fig.~\ref{ResponsesFeP} normalized at $cos^{2}(\Theta)=1$. Right panel: Difference of \emph{Normalized matched fraction} at $\Theta=45^{\circ}$ from $\Theta=0^{\circ}$ depending on iron fraction in p/Fe sample.}
  \label{MatchedFraction}
 \end{figure}

The different EM and $\mu$ detector response to iron nuclei and protons implies a different event ordering according to the signal size in given bin of $cos^{2}(\Theta)$ for mixed composition scenarios. This difference varies with zenith angle (left panel of Fig.~\ref{ResponsesFeP}). Assuming the same exposure for both detectors, we defined \emph{Matched Fraction} being the relative number of the same events present in both sets of \emph{N} events with largest values of signal ($S_{EM}$ or $S_{\mu}$) for each bin in $cos^{2}(\Theta)$. Value of \emph{N} was chosen so that the $N^{th}$ largest signal corresponded to the energy $\sim$10~EeV. Different compositions of primaries were used to estimate \emph{Matched Fraction} as a function of $\Theta$ (right panel of Fig.~\ref{ResponsesFeP}). To exclude the combined effect of different size of fluctuations for iron nuclei and protons we used the \emph{Matched Fraction} normalized to $cos^{2}(\Theta)=1$ (left panel of Fig.~\ref{MatchedFraction}) to study the sensitivity to the mass composition. The size of decrease of \emph{Normalized Matched fraction} with zenith angle seems to be proportional to the amount of admixture. Fitted quadratic function was chosen to describe this behaviour. In the right panel of Fig.~\ref{MatchedFraction} the difference between the \emph{Normalized Matched Fraction} at $cos^{2}(\Theta)=1$ and $cos^{2}(\Theta)=0.5$ is plotted vs. iron nuclei fraction to better visualize the mentioned effect.

\section{Conclusions}
We considered a single observatory consisting of independent EM and $\mu$ surface detector arrays. Relative response to primary protons with respect to iron nuclei showed different zenith angle dependence for EM and $\mu$ detectors. $\Delta$~\emph{Matched fraction} as defined in the paper could be used to distinguish between pure and mixed composition scenarios when CIC method is used to calculate the two attenuation curves at such a combined observatory.

\vspace*{0.5cm}
\footnotesize{{\bf Acknowledgment:~}{This work is funded by Ministry of Education, Youth and Sports of the Czech Republic under the project LG13007.}}

%\bibliographystyle{ws-procs9x6}
%\bibliography{ws-pro-sample}

\begin{thebibliography}{}

\bibitem{TA:enespec} %T.~Abu-Zayyad, R.~Aida, M.~Allen, R.~Anderson, R.~Azuma, E.~Barcikowski, J.~W.~Belz and D.~R.~Bergman {\it et al.},
T.~Abu-Zayyad {\it et al.},
  %``The Cosmic Ray Energy Spectrum Observed with the Surface Detector of the Telescope Array Experiment,''
  Astrophys.\ J.\  768 (2013) L1.% doi:10.1088/2041-8205/768/1/L1.
  %[arXiv:1205.5067 [astro-ph.HE]].
  %%CITATION = ARXIV:1205.5067;%%
  %19 citations counted in INSPIRE as of 21 May 2013

\bibitem{Hersil:cicmet}
%  J.~Hersil, I.~Escobar, D.~Scott, G.~Clark and S.~Olbert,
  J.~Hersil {\it et al.},
  %``Observations of Extensive Air Showers near the Maximum of Their Longitudinal Development,''
  Phys.\ Rev.\ Lett.\  6 (1961) 22-23.% doi:10.1103/PhysRevLett.6.22.
  %%CITATION = PRLTA,6,22;%%
  %21 citations counted in INSPIRE as of 21 May 2013

\bibitem{AUGER:enespec}
  J.~Abraham {\it et al.},  [Pierre Auger Collaboration],
  %``Observation of the suppression of the flux of cosmic rays above $4\times 10^{19}$eV,''
  Phys.\ Rev.\ Lett.\ 101 (2008) 061101.% doi:10.1103/PhysRevLett.101.061101.

\bibitem{CORSIKA}
%  D.~Heck, G.~Schatz, T.~Thouw, J.~Knapp and J.~N.~Capdevielle,
  D.~Heck {\it et al.},
  %``CORSIKA: A Monte Carlo code to simulate extensive air showers,''
  Forschungszentrum Karlsruhe Report FZKA 6019 (1998), 90 pages.
 %FZKA-6019.
  %%CITATION = FZKA-6019;%%
  %29 citations counted in INSPIRE as of 21 May 2013

\bibitem{ICRC} J.~Vicha et al., arXiv:1310.0330 [astro-ph.HE], \emph{Proceedings of ICRC~2013}.

\bibitem{QGSJETII04} 
S.S. Ostapchenko, Phys. Rev. D83 (2011) 014018.

\bibitem{epos} K.~Werner, F.~M.~Liu and T.~Pierog,
               {\it Phys.\ Rev.\  C }  74 (2006) 044902.
 %              doi: 10.1103/PhysRevC.74.044902.

\bibitem{epos2} T. Pierog and K. Werner, {\it Nucl. Phys. }(Proc. Suppl.) 
               196 (2009) 102.
%               doi: 10.1016/j.nuclphysbps.2009.09.017.

\bibitem{FLUKA}
%  A.~Ferrari, P.~R.~Sala, A.~Fasso and J.~Ranft,
  A.~Ferrari {\it et al.},
  %``FLUKA: A multi-particle transport code (Program version 2005),''
  CERN-2005-010.
  %%CITATION = CERN-2005-010;%%
  %122 citations counted in INSPIRE as of 22 May 2013

\end{thebibliography}

\end{document}